\documentclass{natureprintstyle}
\usepackage{graphicx}
\usepackage{amssymb,hyperref}

\date{August 7, 2014 \\ 
      \sf{\small{authors' version, accepted to \emph{Nature}; final version
                 available at \url{http://dx.doi.org/10.1038/nature13615}}}}
\usepackage{multibib}
\newcites{meth}{\mbox{ }}

\bibliographystylemeth{naturemag-swj}

\usepackage{color}
\definecolor{darkgreen}{rgb}{0,0.5,0}
\definecolor{purple}{rgb}{0.8,0,0.8}
\definecolor{darkblue}{rgb}{0,0,0.6}
\definecolor{darkred}{rgb}{0.6,0,0}
\definecolor{orange}{rgb}{1.0,0.5,0}
\definecolor{gold}{rgb}{0.7,0.5,0}
\newcommand{\kibitz}[2]{\ifnum\Comments=1\textcolor{#1}{#2}\fi}

\newcount\Comments  
\newcommand{\sun}{\odot}

\newcommand{\about}{$\approx\!$~}
\newcommand\arcdeg{\mbox{$^\circ$}}%
\newcommand\arcmin{\mbox{$^\prime$}}%
\newcommand\arcsec{\mbox{$^{\prime\prime}$}}%
\newcommand\fs{\mbox{$\,.\!\!^{\sf s}$}}%
\newcommand\fsrm{\mbox{$\,.\!\!^{\textrm{s}}$}}%
\newcommand\farcs{\mbox{$.\!\!^{\prime\prime}$}}

\title{A luminous, blue progenitor system for a type-Iax supernova}

\author{Curtis~McCully$^1$, 
Saurabh~W.~Jha$^1$, 
Ryan~J.~Foley$^{2,3}$, 
Lars~Bildsten$^{4,5}$,\\
Wen-fai~Fong$^6$,
Robert~P.~Kirshner$^6$,
G.~H.~Marion$^{7,6}$,
Adam~G.~Riess$^{8,9}$,\\
\& Maximilian~D.~Stritzinger$^{10}$
}

\begin{document}

\maketitle

\begin{affiliations}
\item Department of Physics and Astronomy, Rutgers, the State
  University of New Jersey, 136 Frelinghuysen Road, Piscataway, New
  Jersey 08854, USA.
\item Astronomy Department, University of Illinois at
  Urbana-Champaign, 1002 W.\ Green Street, Urbana, Illinois 61801, USA.
\item Department of Physics, University of Illinois at
  Urbana-Champaign, 1110 W.\ Green Street, Urbana, Illinois 61801, USA.
\item Department of Physics, University of California, Santa Barbara,
  California 93106, USA. 
\item Kavli Institute for Theoretical Physics, University of
  California, Santa Barbara, California 93106, USA.
\item Harvard-Smithsonian Center for Astrophysics, 60 Garden Street,
  Cambridge, Massachusetts 02138, USA.
\item Department of Astronomy, University of Texas at Austin, Austin,
  Texas 78712, USA.
\item Department of Physics and Astronomy, Johns Hopkins University,
  3400 North Charles Street, Baltimore, Maryland 21218, USA.
\item Space Telescope Science Institute, 3700 San Martin Drive,
  Baltimore, Maryland 21218, USA.
\item Department of Physics and Astronomy, Aarhus University, Ny
  Munkegade 120, DK-8000 Aarhus C, Denmark.
\end{affiliations}

\begin{abstract}
Type-Iax supernovae (SN~Iax) are stellar explosions that are
spectroscopically similar to some type-Ia supernovae (SN~Ia) at
maximum light, except with lower ejecta velocities\cite{Foley13,
  Li03}. They are also distinguished by lower luminosities. At late
times, their spectroscopic properties diverge from other
SN\cite{Jha06,Phillips07,Sahu08,McCully14}, but their composition
(dominated by iron-group and intermediate-mass elements\cite{Foley13,
  Stritzinger14}) suggests a physical connection to normal
SN~Ia. These are not rare; SN~Iax occur at a rate between 5 and 30\%
of the normal SN~Ia rate\cite{Foley13}. The leading models for SN~Iax
are thermonuclear explosions of accreting carbon-oxygen white dwarfs
(C/O WD) that do not completely unbind the
star\cite{Jordan12,Kromer13,Fink14}, implying they are ``less
successful'' cousins of normal SN~Ia, where complete disruption is
observed.  Here we report the detection of the luminous, blue
progenitor system of the type-Iax SN~2012Z in deep pre-explosion
imaging. Its luminosity, colors, environment, and similarity to the
progenitor of the Galactic helium nova V445
Puppis\cite{Kato08,Woudt09,Goranskij10}, suggest that SN~2012Z was the
explosion of a WD accreting from a helium-star companion. Observations
in the next few years, after SN~2012Z has faded, could test this
hypothesis, or alternatively show that this supernova was actually the
explosive death of a massive star\cite{Valenti09,Moriya10}.
\end{abstract}

SN~2012Z was discovered\cite{Cenko12} in the Lick
Observatory Supernova Search on UT 2012-Jan-29.15. It had an optical
spectrum similar to the type-Iax (previously called SN~2002cx-like)
SN~2005hk\cite{Jha06,Phillips07,Sahu08} (see Extended Data
Fig.~\ref{fig:spec}). The similarities between SN~Iax and normal SN~Ia
make understanding SN~Iax progenitors important, especially because no
normal SN~Ia progenitor has been identified.  Like core-collapse SN
(but also slowly-declining, luminous SN~Ia), SN~Iax are found
preferentially in young, star-forming galaxies\cite{Foley09,
  Lyman13}. A single SN~Iax, SN~2008ge, was in a relatively old (S0)
galaxy with no indication of current star formation to deep
limits\cite{Foley10_ge}. Non-detection of the progenitor of SN~2008ge
in \emph{Hubble Space Telescope} (\emph{HST}) pre-explosion imaging
restricts its initial mass $\lesssim$12~M$_{\sun}$, and combined with
the lack of hydrogen or helium in the SN~2008ge spectrum, favours a
white dwarf progenitor\cite{Foley10_ge}. 

Deep observations of NGC~1309, the host galaxy of SN~2012Z, were
obtained with \emph{HST} in 2005--2006 and 2010, serendipitously
including the location of the supernova before its explosion. To
pinpoint the position of SN~2012Z with high precision, we obtained
follow-up \emph{HST} data in 2013. Colour-composite images made from
these observations before and after the supernova are shown in
Fig.~\ref{fig:image}, and photometry of stellar sources in the
pre-explosion images near the supernova location is reported in
Extended Data Table \ref{tab:stars}. We detect a source, called S1,
coincident with the supernova at a formal separation of 0{\farcs}0082
$\pm$ 0{\farcs}0103 (equal to 1.3 $\pm$ 1.6 pc at 33 Mpc, the distance
to NGC~1309\cite{Riess09b,Riess11}). The pre-explosion data reach a
3$\sigma$ limiting magnitude of M$_{\rm V} \approx -$3.5, quite deep
for typical extragalactic SN progenitor searches\cite{Smartt09}, but
certainly the possibility exists that the progenitor system of
SN~2012Z was of lower luminosity and would be undetected in our data
(as has been the case for all normal SN~Ia progenitor searches to
date\cite{Li11}). However, the locations of SN~2012Z and S1 are
identical to within 0.8$\sigma$, and we estimate only a 0.24\%
(2.1\%) probability that a random position near SN~2012Z would be
within 1$\sigma$ (3$\sigma$) of any detected star, making a chance
alignment unlikely (see Methods, and Extended Data
  Fig.~\ref{fig:chance}). We also observe evidence for variability in
S1 (plausible for a pre-supernova system; Extended Data Table
\ref{tab:variability}), at a level exhibited by only 4\% of objects of
similar brightness. We thus conclude there is a high likelihood that
S1 is the progenitor system of SN~2012Z.

The color-magnitude diagram (CMD) presented in Fig.~\ref{fig:cmd}
shows S1 to be luminous and blue, yet in an odd place for a star about
to explode. If its light is dominated by a single star, S1 is
moderately consistent with a \about 18.5~M$_{\sun}$ main-sequence
star\cite{Bertelli09}, an \about 11~M$_\sun$ blue supergiant early in
its evolution off the main sequence, or perhaps a \about
7.5~M$_{\sun}$ (initial mass) blue supergiant later in its evolution
(with core helium-burning in a blue loop, where models are quite
sensitive to metallicity and rotation\cite{Georgy13}). None of these
stars are expected to explode in standard stellar evolution theory,
particularly without any signature of hydrogen in the
supernova\cite{Smartt09}.

The SN~2012Z progenitor system S1 is in a similar region in the CMD to
some Wolf-Rayet stars\cite{Shara13}, highly evolved, massive stars,
that are expected to undergo core collapse and may produce a
supernova. If S1 were a single Wolf-Rayet star, its photometry is most
consistent with the WN subtype and an initial mass \about
30--40~M$_\sun$, thought perhaps to explode with a helium-dominated
outer layer as a SN~Ib\cite{Groh13}, and unlikely to produce the
structure and composition of ejecta seen in SN~Iax\cite{Jha06,
  Foley13, McCully14, Stritzinger14}. Moreover,
isochrones\cite{Bertelli09} fit to the neighbouring stars (Extended
Data Fig.~\ref{fig:isochrone}) yield an age range of \about 10--42
Myr, longer than the 5--8 Myr lifetime of such a massive Wolf-Rayet
star.

S1 may be dominated by accretion luminosity; its brightness in B and V
is not far from the predicted thermal emission of an
Eddington-luminosity Chandrasekhar mass WD (a super-soft source; SSS;
Fig.~\ref{fig:cmd}). However, its V-I and V-H colours are too red for
a SSS model. A composite scenario, with accretion power dominating the
blue flux, and another source providing the redder light (perhaps a
fainter, red donor star) may be plausible.

The leading models of SN~Iax\cite{Jordan12, Kromer13, Fink14} are
based on C/O WD explosions, so S1 may be the companion star to an
accreting WD. Although there are a variety of potential progenitor
systems (including main-sequence and red giant donors, which are
inconsistent with S1 if they dominate the system's luminosity), in
standard scenarios no companion star can have an initial mass greater
than \about 7~M$_{\sun}$; otherwise, there would not be enough time to
form the primary C/O WD that explodes. Thus, the photometry of S1
suggests that if it is the companion to a C/O WD, recent binary mass
transfer must have played a role in its evolution. One model for a
luminous, blue companion star is a relatively massive (\about 2
M$_{\sun}$ when observed) helium star\cite{Iben91, Kato08, Liu10},
formed after binary mass transfer and a common envelope phase (e.g., a
close binary with initial masses \about 7 and 4~M$_\sun$). Although
the model parameter space has not been fully explored, the predicted
region for helium star donors in a binary system with a 1.2~M$_{\sun}$
initial-mass accreting C/O WD\cite{Liu10} in the CMD is shown in
Fig.~\ref{fig:cmd}, and S1 is consistent with being in this
region. The evolutionary timescale for such a model is also
well-matched to the ages of nearby stars (Extended Data
Fig.~\ref{fig:isochrone}).

SN~2012Z and the star S1 have an interesting analogue in our own Milky
Way Galaxy: the helium nova
V445~Puppis\cite{Kato08,Woudt09,Goranskij10}, thought to be the
surface explosion of a near-Chandrasekhar mass helium-accreting
WD. Though S1 is somewhat brighter than the pre-explosion observations
of V445~Pup, their consistent colours, similar variability
amplitude\cite{Goranskij10}, and the physical connection between
V445~Pup and likely SN~Iax progenitors\cite{Jordan12,Kromer13,Fink14}
is highly suggestive. Indeed, two SN~Iax (though not SN~2012Z itself)
have shown evidence for helium in the system\cite{Foley09,Foley13}. In
this model, a low helium accretion rate could lead to a helium nova
(like V445 Pup), whereas a higher mass-transfer rate could result in
stable helium burning on the C/O WD, allowing it to grow in mass
before the supernova. The accretion is expected to begin as the helium
star starts to evolve and grow in radius; indeed, the S1 photometry is
consistent with the evolutionary track of a helium star with a mass
(after losing its hydrogen envelope) of \about 2 M$_{\sun}$, on its
way to becoming a red giant\cite{Kato08}.

Though the scenario of a helium-star donor to an exploding
carbon-oxygen white dwarf is a promising model for the progenitor and
supernova observations, we cannot yet rule out the possibility that S1
is a single star that itself exploded. Fortunately, by late 2015,
SN~2012Z will have faded below the brightness of S1, and \emph{HST}
imaging will allow us to distinguish these models.  Our favoured
interpretation of S1 as the companion star predicts that it will still
be detected (though perhaps modified by the impact of its exploding
neighbour, a reduction in accretion luminosity, or a cessation of
variability).  On the other hand, if S1 has completely disappeared, it
will be a strong challenge to models of SN~Iax, and perhaps
significantly blur the line between thermonuclear white-dwarf
supernovae and massive-star core-collapse supernovae, with important
impacts to our understanding of stellar evolution and chemical
enrichment.

\bibliography{progenitor-12Z}

\begin{addendum}

\item[Acknowledgments]
Supernova 2012Z was discovered just weeks after the passing of our
dear friend and colleague, Weidong Li, whose work on the Lick
Observatory Supernova Search, SN 2002cx-like supernovae, and Hubble
Space Telescope observations of SN progenitors, continues to inspire
us. That those three foci of Weidong's research converge here in
this paper makes our hearts glad, and we dedicate this paper to his
memory.

We thank the SH$_0$ES team for assistance with data from \emph{HST}
program GO-12880, E.~Bertin for the development of the STIFF software
to produce color images, and A.~Dolphin for software and guidance in
photometry.  This research at Rutgers University was supported through
NASA/{\it HST} grant GO-12913.01, and National Science Foundation
(NSF) CAREER award AST-0847157 to S.W.J. NASA/\emph{HST} grant
GO-12999.01 to R.J.F. supported this work at the University of
Illinois. At UC Santa Barbara, this work was supported by NSF grants
PHY 11-25915 and AST 11-09174 to L.B. The Danish Agency for Science,
Technology, and Innovation supported M.D.S. through a Sapere Aude
Level 2 grant.

Support for \emph{HST} programs GO-12913 and GO-12999 was provided by
NASA through a grant from the Space Telescope Science Institute, which
is operated by the Association of Universities for Research in
Astronomy, Incorporated, under NASA contract NAS5-26555.

\item[Author Contributions] 
C.M., S.W.J., and R.J.F.~performed the data analysis and were chiefly
responsible for preparing the manuscript and figures. W.F., R.P.K.,
G.H.M., and A.G.R. assisted in developing the proposal to obtain
\emph{HST} observations, including acquiring supporting ground-based
photometry and spectroscopy. L.B.~provided insight into models for
progenitor systems. M.D.S.~analyzed ground-based photometry and
spectroscopy of the supernova, used as input for this paper.  All
authors contributed to extensive discussions about, and edits to, the
paper draft.

\item[Competing Interests] 
The authors declare that they have no competing financial interests.

\item[Correspondence] 
Correspondence and requests for materials should be addressed to
S.W.J.~(saurabh@physics.rutgers.edu).

\end{addendum}

\section*{Methods}

\mbox{ }

\vskip 0.05in 
\noindent 

\noindent \textbf{Observations and reduction.} SN~2012Z provides a unique
opportunity to search for a SN~Iax progenitor, because of deep,
pre-explosion \emph{HST} observations. Its face-on spiral host galaxy
NGC~1309 was also the site of the nearby, normal type-Ia SN~2002fk,
and as such was targeted in 2005 and 2006 with the \emph{HST} Advanced
Camera for Surveys (ACS) and in 2010 with the \emph{HST} Wide-Field
Camera 3 (WFC3) optical (UVIS) and infrared (IR) channels, to observe
Cepheid variable stars in order to anchor the SN~Ia distance scale
(\emph{HST} programs GO-10497, GO-10802, and GO-11570, PI: A.~Riess;
GO/DD-10711, PI: K.~Noll). To measure the location of SN~2012Z with
high precision, we used \emph{HST} WFC3/UVIS images of NGC~1309
(fortuitously including SN~2012Z) taken on UT 2013-Jan-04 (program
GO-12880; PI: A.~Riess), as well as targeted \emph{HST} WFC3/UVIS
images of SN~2012Z taken on UT 2013-Jun-30 (GO-12913; PI: S.~Jha).

\begin{figure*}
\hskip -0.35in 
\includegraphics[width=1.12\textwidth]{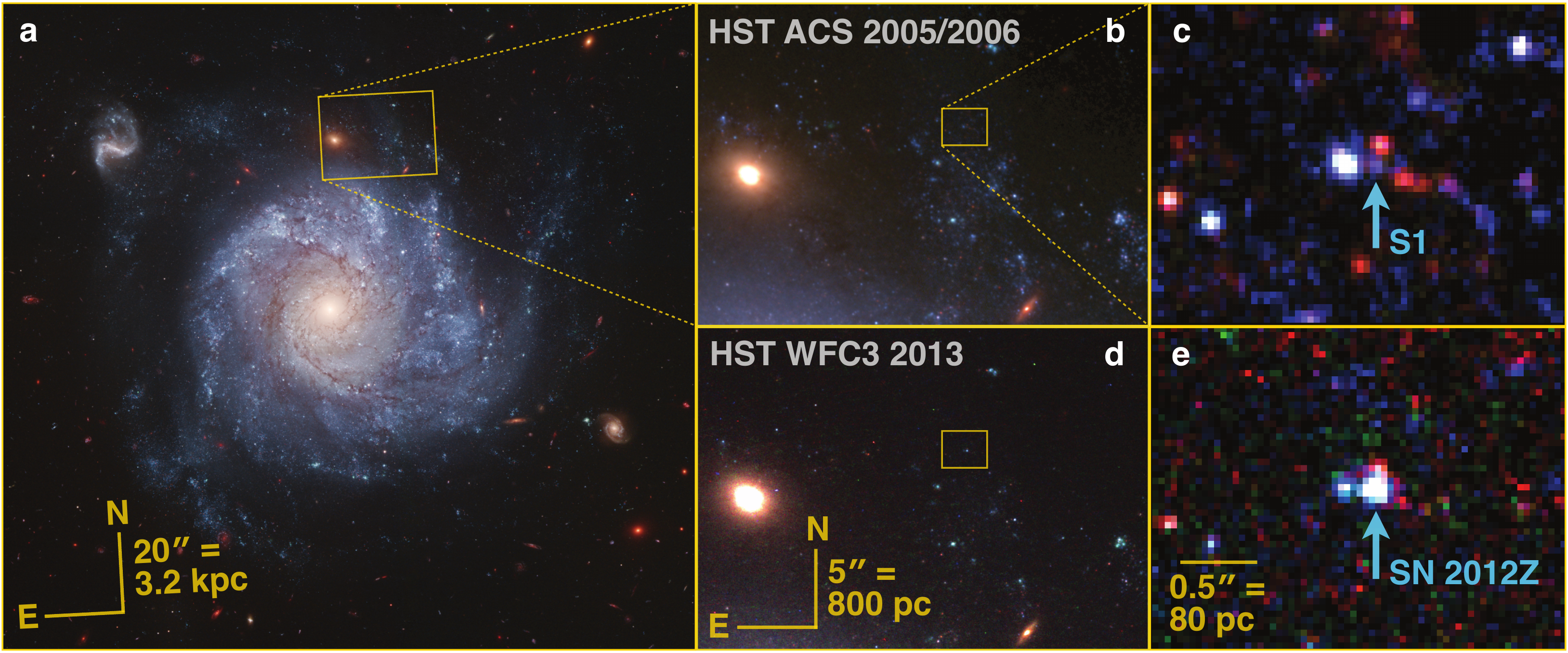}
\vskip -0.1in
\caption{\textbf{\emph{Hubble Space Telescope} colour images before and
    after supernova~2012Z.} Panel \textbf{a} shows the Hubble Heritage
  image of NGC~1309 (\url{http://heritage.stsci.edu/2006/07}), with
  panels \textbf{b} and \textbf{c} zooming in on the progenitor system
  S1 in the deep, pre-explosion data. Lower panels \textbf{d} and \textbf{e}
  show the shallower post-explosion images of SN~2012Z on the same
  scale as the panels above. 
\label{fig:image}}
\end{figure*}

The 2005--2006 ACS images include 2 visits totaling 9600s of exposure
time in the F435W filter (similar to Johnson $B$), 14 visits for a
total exposure of 61760 sec in F555W (close to Johnson $V$) and 5
visits for 24000s in F814W (analogous to Cousins $I$). These are
showing the blue, green, and red channels, respectively of panels a--c
in Fig.~\ref{fig:image}. We re-reduced the archived data, combining
the multiple exposures (including sub-sampling and cosmic ray
rejection) using the AstroDrizzle software from the DrizzlePac
package\citemeth{Gonzaga12}, with the results shown in panels b and c
of Fig.~\ref{fig:image}. The bottom right panels d and e of
Fig.~\ref{fig:image} show our combined \emph{HST} WFC3/UVIS F555W
(blue; 1836 sec) + F625W (green; 562 sec) + F814W (red; 1836 sec)
images from January and June 2013, with SN~2012Z visible.

We used the DrizzlePac TweakReg routine to register all of the
individual flatfielded (``{\tt flt}'') frames to the WFC3/UVIS F555W
image taken on UT 2013-Jan-04. The typical root-mean-square (rms)
residual of individual stars from the relative astrometric solution
was 0\farcs009, corresponding to 0.18 pixels in ACS and 0.23 pixels in
WFC3/UVIS. We drizzle the ACS images to the native scale of UVIS,
0\farcs04 per pixel (20\% smaller than the native 0\farcs05 ACS
pixels) and subsample the ACS point-spread function (psf)
correspondingly with a pixel fraction parameter of 0.8.

\noindent \textbf{Photometry and astrometry.} In Extended Data Table
\ref{tab:stars}, we present photometry of sources in the region based
on the \emph{HST} ACS images from 2005--2006 (F435W, F555W, and
F814W), as well as WFC3/IR F160W data (6991 sec of total exposure
time) from 2010. The stars are sorted by their proximity to the SN
position, and their astrometry is referenced to SDSS images of the
field, with an absolute astrometric uncertainty of 0\farcs080 (but
this is irrelevant for the much more precise relative astrometry of
SN~2012Z and S1).  We photometered the \emph{HST} images using the
psf-fitting software DolPhot, an extension of
HSTPhot\citemeth{Dolphin00}. We combined individual flt frames taken
during the same \emph{HST} visit at the same position, and then used
DolPhot to measure photometry using recommended parameters for ACS and
WFC3.

The WFC3/UVIS images of SN~2012Z from January and June 2013 provide a
precise position for the supernova of R.A. =
3$^{\textrm{h}}$22$^{\textrm{m}}$05\fsrm39641, Decl. = $-$15\arcdeg
23\arcmin 14{\farcs}9390 (J2000) with a registration uncertainty of
0\farcs0090 (plus an absolute astrometric uncertainty of 0\farcs08,
irrelevant to the relative astrometry). As shown in
Fig.~\ref{fig:image} we detect a stellar source (called S1) in the
pre-explosion images coincident with the position of the SN with a
formal separation, including centroid uncertainties, of 0{\farcs}0082
$\pm$ 0{\farcs}0103, indicating an excellent match and strong evidence
for S1 being the progenitor system of SN~2012Z.

\noindent \textbf{Chance alignment probability.} To estimate the
probability of a chance alignment, we use the observed density of
sources detected with signal-to-noise ratio S/N $>$ 3.0 (in any
filter) and S/N $>$ 3.5 (in all bands combined, via DolPhot) in a 200
$\times$ 200~pixel (8$\arcsec$ $\times$ 8$\arcsec$) box centered on
SN~2012Z (452 sources), and find only a 0.24\% (2.1\%) chance that a
random position would be consistent with a detected star at 1$\sigma$
(3$\sigma$). Moreover, only 171 of these stars are as bright as S1, so
\emph{a posteriori} there was only a 0.09\% (0.80\%) chance of a
1$\sigma$ (3$\sigma$) alignment with such a bright object. As shown in
Extended Data Fig.~\ref{fig:chance}, these results are not especially
sensitive to the size of the region used to estimate the density of
sources, at least down to 50 $\times$ 50~pixels (2$\arcsec$ $\times$
2$\arcsec$) around SN~2012Z. Nearer than this, the density of sources
increases by a factor of a few, though with substantially larger
uncertainty given the low number of sources (including S1 itself). We
base our fiducial chance alignment probability on the larger region
where the density of sources stabilizes with good statistics, but our
qualitative results do not depend on this choice.

Given the surface brightness of NGC~1309, we crudely estimate \about
160 M$_\sun$ in stars projected within an area corresponding to our
1$\sigma$ error circle (\about 8 pc$^2$). These should be roughly
uniformly distributed throughout this small region, so our chance
alignment probability should accurately quantify the probability that
SN~2012Z originated from an undetected progenitor that was only
coincidentally near a detected source like S1.

The pre-explosion data for SN~2012Z are the deepest ever for a SN~Iax,
reaching M$_{\rm V} \approx -$3.5; the next best limits come from
SN~2008ge\cite{Foley10_ge}, which yielded no progenitor detection down
to M$_{\rm V} \approx -$7. The star S1, at M$_{\rm V} \approx -$5.3,
would not have been detected in any previous search for SN~Iax
progenitors. This implies that our chance alignment probability
calculation can be taken at face value; there have not been previous,
unsuccessful ``trials'' that would reduce the unlikelihood of a chance
coincidence. Viewed in the context of progenitor searches for normal
SN Ia, in only two cases: SN~2011fe\cite{Li11} and
SN~2014J\cite{Kelly14}, would a star of the luminosity of S1 have been
clearly detected in pre-explosion data. Other normal SN Ia, like
SN~2006dd\cite{Maoz08}, have progenitor detection limits in
pre-explosion observations just near, or above, the luminosity of S1.

\noindent \textbf{Variability of S1.} NGC 1309 was imaged over 14 epochs in
F555W with ACS before SN 2012Z exploded. Examining these individually,
we find some evidence for variability in S1; the photometry is
presented in Extended Data Table \ref{tab:variability}. Formally,
these data rule out the null hypothesis of no variability at 99.95\%
(3.5$\sigma$), with $\chi^2 =$ 36.658 in 13 degrees of
freedom. However, most of the signal is driven by one data point (MJD
53600.0; a 4.2$\sigma$ outlier); excluding this data point (though we
find no independent reason to do so) reduces the significance of the
variability to just 91.1\% (1.7$\sigma$). To empirically assess the
likelihood of variability, we looked to see how often stars with the
same brightness as S1 (within 0.5 mag) doubled in brightness relative
to their median like S1 in one or more of the 14 epochs; we find that
just 4\% do so (4 of the nearest 100, 9 of the nearest 200, and 20 of
nearest 500 such stars). As variability might be expected in a SN
progenitor before explosion (from non-uniform accretion, for example),
combining this with the chance alignment probability strengthens the
identification of S1 as the progenitor system of SN~2012Z. It also
disfavours the possibility that S1 is a compact, unresolved star
cluster.

\begin{figure*}
\begin{center}
\includegraphics[width=\textwidth]{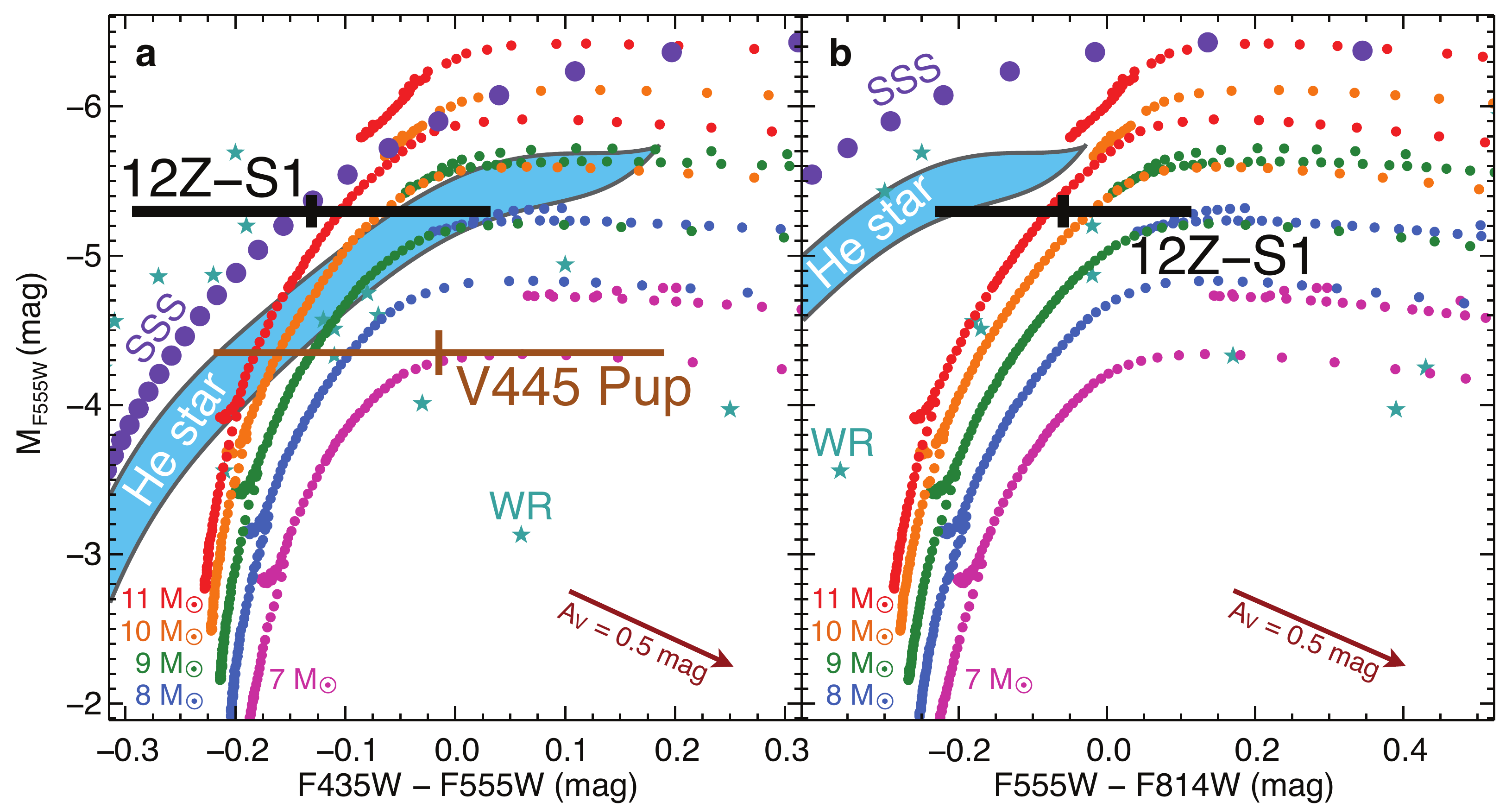}
\end{center}
\vskip -0.2in
\caption{\textbf{Colour-magnitude diagrams of the SN~2012Z progenitor
    system S1 and comparison models.} Panel \textbf{a} presents the
  F435W-F555W colour (roughly B-V), while panel \textbf{b} shows
  F555W-F814W (V-I), both plotted against the F555W (V) absolute
  magnitude. The black and brown crosses represent the progenitor
  systems for SN~2012Z and V445~Puppis\cite{Woudt09}, respectively,
  with 1$\sigma$ photometric uncertainties. Other comparisons plotted
  include evolutionary tracks\cite{Bertelli09} for single stars
  (coloured dotted curves), thermal models for Eddington-luminosity
  super-soft sources (SSS; purple dots), candidate Wolf-Rayet
  stars\cite{Shara13} (blue-gray stars), and models for helium-star
  donors to 1.2~M$_{\sun}$ initial mass C/O WDs\cite{Liu10} (shaded
  blue regions).
\label{fig:cmd}}
\end{figure*}

\noindent \textbf{Properties of S1 and nearby stars.} In
Fig.~\ref{fig:cmd} we present colour-magnitude diagrams of S1 and
other objects for comparison. In the figure, S1 has been corrected for
Milky Way reddening (E(B-V)$_{\rm{MW}}$ = 0.035 mag, corresponding to
A$_{\rm{F435W}}$ = 0.14 mag, A$_{\rm{F555W}}$ = 0.11 mag,
A$_{\rm{F814W}}$ = 0.06 mag), and host reddening (E(B-V)$_{\rm{host}}$
= 0.07 mag; A$_{\rm{F435W}}$ = 0.28 mag, A$_{\rm{F555W}}$ = 0.22 mag,
A$_{\rm{F814W}}$ = 0.12 mag) based on narrow, interstellar absorption
lines in high-resolution spectroscopy of SN~2012Z
itself\cite{Stritzinger14b}. This low extinction is consistent with
the photometry and spectroscopy of SN~2012Z, as well as its location
in the outskirts of a face-on spiral host. For the potential Galactic
analogue V445~Puppis progenitor, we correct its photometry for
Galactic and circumstellar reddening\cite{Woudt09}.

Fig.~\ref{fig:cmd} also shows stellar evolution
tracks\cite{Bertelli09} for stars with initial masses of 7, 8, 9, 10,
and 11 M$_{\sun}$, adopting a metallicity of 0.87 solar, based on the
H II region metallicity gradient\cite{Riess09b} for NGC 1309
interpolated to the SN radial location. The Eddington-luminosity
accreting Chandrasekhar-mass white dwarfs are shown as the large
purple dots, with each subsequent dot representing a change in
temperature of 1000~K.  These ``super-soft'' sources are fainter in
F555W for higher temperatures (and bluer F435W$-$F555W colours) as the
fixed (Eddington-limited) bolometric luminosity emerges in the
ultraviolet for hotter systems. The shaded blue region represents the
range of helium-star donors for C/O WD SN progenitor models starting
with a 1.2~M$_{\sun}$ white dwarf\cite{Liu10}. We converted the model
temperatures and luminosities to our observed bands assuming a
blackbody spectrum. The expected temperature and luminosity for this
class of models is expected to vary with white dwarf mass, and
therefore, we regard this region as approximate; its shape, size, and
location is subject to change.

S1 is inconsistent with all confirmed progenitors of core-collapse SN
(exclusively SN~II), which are mostly red
supergiants\cite{Smartt09}. The blue supergiant progenitor of, e.g.,
SN~1987A, was significantly more luminous and likely more massive than
S1\citemeth{Arnett89}. However, we caution that our theoretical
expectations for massive stars could be modified if S1 is in a close
binary system where mass transfer has occurred.

The stars detected in the vicinity of S1 provide clues to the nature
of recent star formation in the region. They include red supergiants
(like S2 and S3 from Table \ref{tab:stars}) as well as objects bluer
and more luminous than S1 (like S5). We show CMDs including these
stars in Extended Data Fig.~\ref{fig:isochrone}; the stars plotted
have a signal-to-noise ratio S/N $>$ 3.5 in the displayed filters, and
were required to be no closer than 3 pixels to a brighter source to
avoid photometric uncertainties from crowding. Model
isochrones\cite{Bertelli09} imply that these stars span an age range
of \about 10--42 Myr. These tracks favour an initial mass for S1 of
\about 7--8~M$_\sun$ (neglecting mass transfer) if it is roughly
coeval with its neighbours. In other words, if S1 were a
30--40~M$_\sun$ initial-mass Wolf-Rayet star with a predicted lifetime
of only 5--8 Myr\cite{Groh13}, it would be the youngest star in the
region.

\bibliographymeth{meth}

\setcounter{figure}{0}
\renewcommand{\figurename}{Extended Data Figure}
\renewcommand{\tablename}{Extended Data Table}

\begin{figure*}
\begin{center}
\includegraphics[width=5.8in]{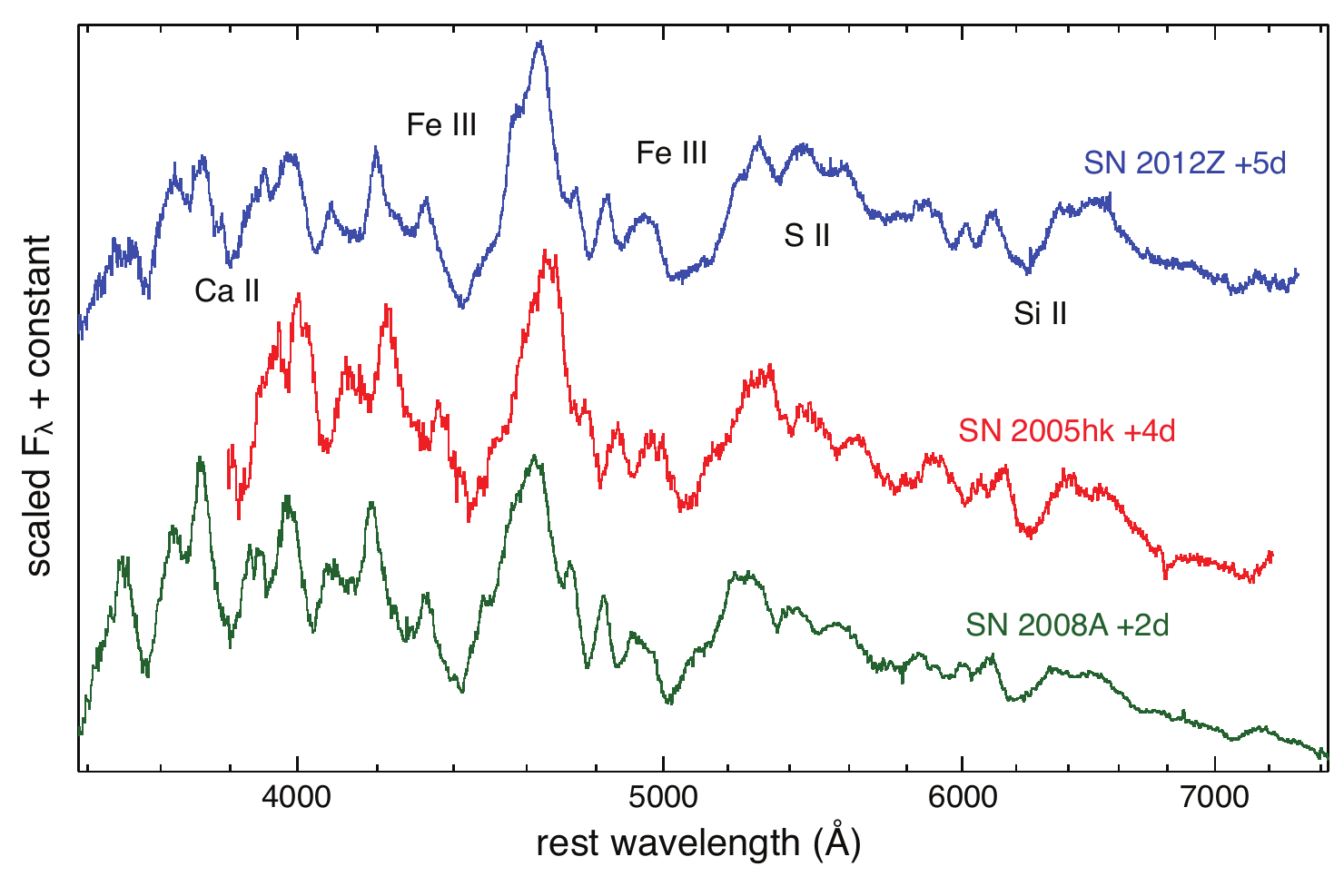}
\end{center}
\vskip -0.2in
\caption{\textbf{Spectra of SN~Iax near maximum light.}
  SN~2012Z\cite{Foley13} is similar to SN~2005hk\cite{Phillips07} and
  SN~2008A\cite{McCully14}, all classified as SN~Iax. The SN~2012Z
  spectrum was taken on UT 2012-02-16 with the Whipple Observatory
  1.5m telescope (+FAST spectrograph) with a total exposure time of
  1800 sec. Each spectrum is labelled by its rest-frame phase past
  \emph{B} maximum light. Prominent features due to intermediate-mass
  and iron group elements are indicated; these features are also
  observed in luminous, slowly-declining SN~Ia spectra at maximum
  light, though with higher expansion velocities.\label{fig:spec}}
\end{figure*}

\begin{figure*}
\begin{center}
\includegraphics[width=5.8in]{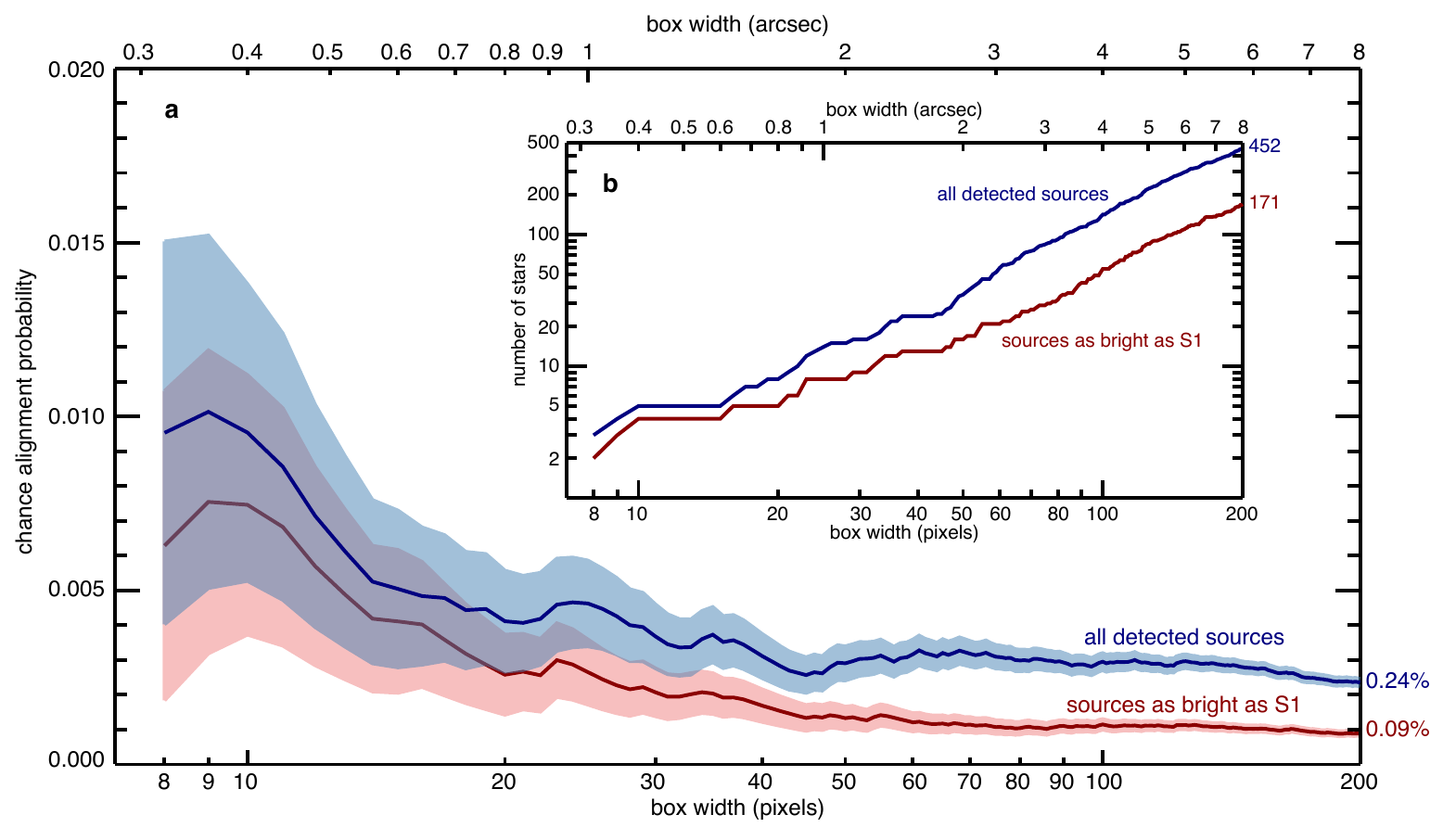}
\end{center}
\vskip -0.2in
\caption{\textbf{Chance alignment probability calculation.}  Panel
  \textbf{a} shows the chance alignment probability between a random
  position and a detected source within a square box of the given
  width centered at the position of SN~2012Z. This calculation is
  based on a 1$\sigma$ position coincidence (0\farcs0103 = 0.2575
  WFC3/UVIS pixels); a 3$\sigma$ position coincidence gives
  probabilities approximately 9 times higher. The offset between
  SN~2012Z and S1 is 0.8$\sigma$. The shaded regions show a naive
  Poisson-like uncertainty estimate with a fractional error given by
  $1/\sqrt{N_{\sf sources}}$. Panel \textbf{b} shows the number of
  detected sources (including S1) as a function of the box size. The
  lines show results for all stellar sources ($>$3$\sigma$ detection
  in any band; blue) and just those as bright as S1
  (red). \label{fig:chance}}
\end{figure*}

\begin{figure*}
\begin{center}
\includegraphics[width=6in]{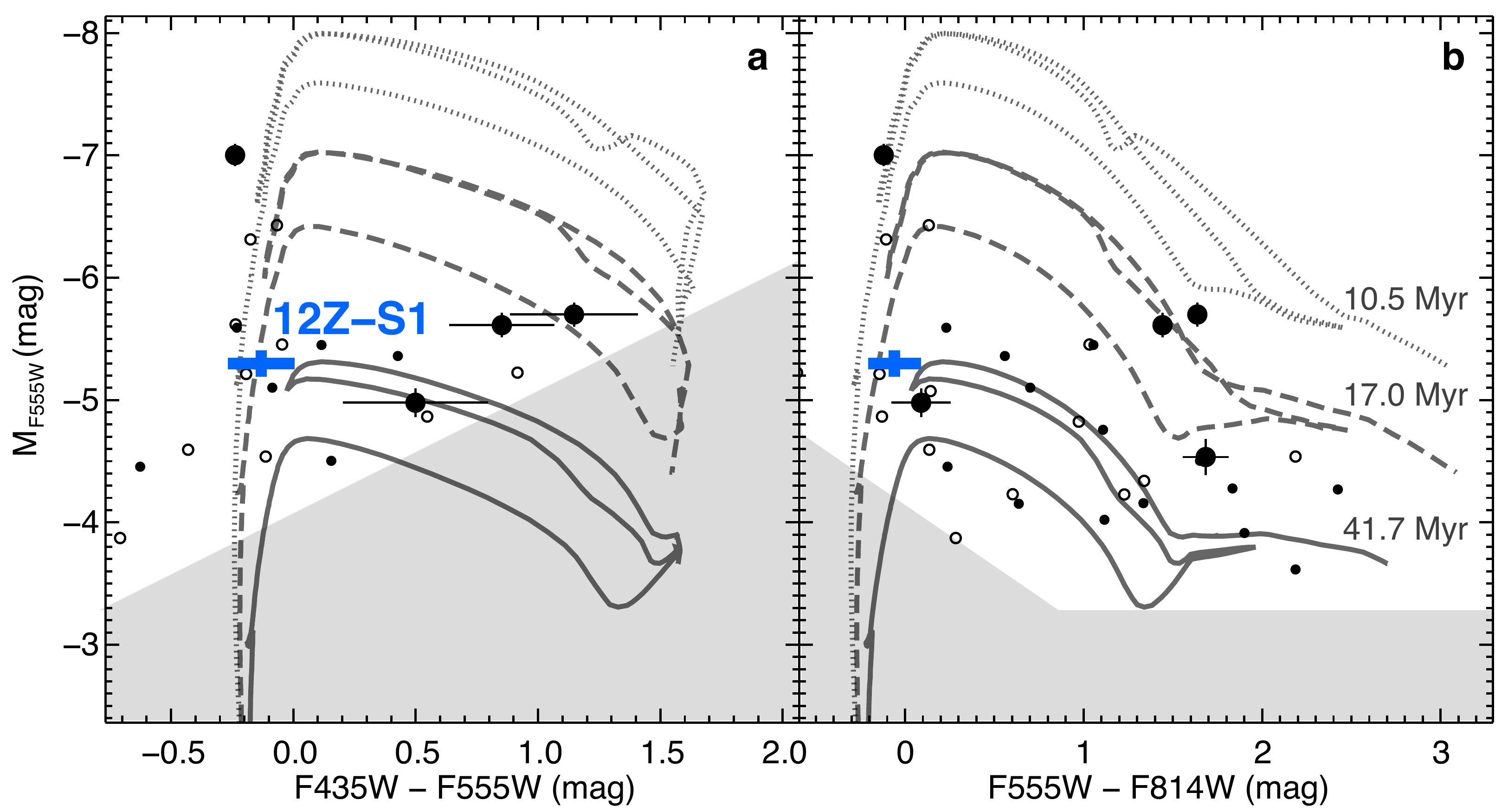}
\end{center}
\vskip -0.2in
\caption{\textbf{Stars in the neighbourhood of the SN~2012Z
    progenitor.}  The SN~2012Z progenitor system S1 (blue) is shown
  along with nearby stars (all with 1$\sigma$ photometric
  uncertainties), with three isochrones\cite{Bertelli09} with ages of
  10.5 (dotted), 17.0 (dashed), and 41.7 Myr (solid).  Panel
  \textbf{a} presents the F435W-F555W colour (roughly B-V), while
  panel \textbf{b} shows F555W-F814W (V-I), both plotted against the
  F555W (V) absolute magnitude. The large, filled circles correspond
  to stars within 10 WFC3/UVIS pixels (0\farcs4) of the SN location,
  small filled circles are within 20 pixels (0\farcs8), and the small
  open circles are within 30 pixels (1\farcs2). Objects in the gray
  shaded regions would not be detected given the depth of the combined
  images.
\label{fig:isochrone}}
\end{figure*}

\clearpage

\begin{table*}
\centering
\caption{\textbf{Pre-explosion photometry of SN~2012Z progenitor
    system S1
    and nearby stars.} The 1$\sigma$ photometric uncertainty is given
  in parentheses and non-detections are listed with 3$\sigma$
  upper-limits.  No corrections for Galactic or host extinction are
  made here. \label{tab:stars}}
\bigskip
\small
\footnotesize
\sffamily
\begin{tabular}{c c c c c c c}
\hline
Star & R.A. (J2000) & Decl. (J2000) & F435W (mag) & F555W (mag)  &
F814W (mag)  & F160W (mag) \\
\hline
S1   & 3$^{\sf h}$22$^{\sf m}$05\fs39591 & $-$15\arcdeg 23\arcmin 14\farcs9350 & 27.589~(0.122) & 27.622~(0.060) & 27.532~(0.135) & 26.443~(0.321) \\
S2   & 3$^{\sf h}$22$^{\sf m}$05\fs40280 & $-$15\arcdeg 23\arcmin 14\farcs9402 & $>$29.221      & 28.551~(0.116) & 27.463~(0.093) & 26.032~(0.238) \\
S3   & 3$^{\sf h}$22$^{\sf m}$05\fs39483 & $-$15\arcdeg 23\arcmin 14\farcs8102 & 28.466~(0.258) & 27.221~(0.041) & 25.435~(0.022) & 23.699~(0.027) \\
S4   & 3$^{\sf h}$22$^{\sf m}$05\fs38460 & $-$15\arcdeg 23\arcmin 15\farcs0314 & 28.258~(0.211) & 27.308~(0.046) & 25.717~(0.028) & 23.887~(0.033) \\
S5   & 3$^{\sf h}$22$^{\sf m}$05\fs41013 & $-$15\arcdeg 23\arcmin 14\farcs9362 & 25.778~(0.029) & 25.918~(0.014) & 25.888~(0.030) & 25.435~(0.136) \\
S6   & 3$^{\sf h}$22$^{\sf m}$05\fs37564 & $-$15\arcdeg 23\arcmin 15\farcs0702 & $>$28.807      & 28.386~(0.116) & 26.553~(0.055) & 24.986~(0.085) \\
S7   & 3$^{\sf h}$22$^{\sf m}$05\fs41970 & $-$15\arcdeg 23\arcmin 14\farcs8454 & 28.539~(0.286) & 27.942~(0.078) & 27.703~(0.146) & 26.436~(0.322) \\
S8   & 3$^{\sf h}$22$^{\sf m}$05\fs37049 & $-$15\arcdeg 23\arcmin 14\farcs6690 & $>$28.847      & 28.763~(0.164) & 27.279~(0.101) & 25.032~(0.092) \\
S9   & 3$^{\sf h}$22$^{\sf m}$05\fs36477 & $-$15\arcdeg 23\arcmin 15\farcs0634 & 27.683~(0.127) & 27.470~(0.050) & 26.266~(0.042) & 24.753~(0.069) \\
S10  & 3$^{\sf h}$22$^{\sf m}$05\fs39605 & $-$15\arcdeg 23\arcmin 15\farcs4062 & $>$28.912      & 29.305~(0.273) & 26.969~(0.077) & 25.214~(0.102) \\
S11  & 3$^{\sf h}$22$^{\sf m}$05\fs37904 & $-$15\arcdeg 23\arcmin 14\farcs5218 & 27.196~(0.083) & 27.329~(0.045) & 26.949~(0.076) & $>$26.614 \\
S12  & 3$^{\sf h}$22$^{\sf m}$05\fs37556 & $-$15\arcdeg 23\arcmin 14\farcs4202 & $>$28.897      & 28.302~(0.109) & 27.095~(0.085) & 25.381~(0.131) \\
S13  & 3$^{\sf h}$22$^{\sf m}$05\fs36527 & $-$15\arcdeg 23\arcmin 14\farcs5054 & 28.085~(0.181) & 27.560~(0.055) & 26.852~(0.069) & 25.740~(0.171) \\
S14  & 3$^{\sf h}$22$^{\sf m}$05\fs40432 & $-$15\arcdeg 23\arcmin 15\farcs5778 & $>$29.360      & 28.651~(0.113) & 26.078~(0.026) & 24.180~(0.042) \\
S15  & 3$^{\sf h}$22$^{\sf m}$05\fs43040 & $-$15\arcdeg 23\arcmin 14\farcs4706 & 27.937~(0.164) & 28.465~(0.122) & 28.078~(0.208) & $>$26.512 \\
S16  & 3$^{\sf h}$22$^{\sf m}$05\fs42376 & $-$15\arcdeg 23\arcmin 14\farcs3726 & $>$28.765      & 28.163~(0.092) & 26.905~(0.072) & 25.126~(0.094) \\
S17  & 3$^{\sf h}$22$^{\sf m}$05\fs38194 & $-$15\arcdeg 23\arcmin 14\farcs2494 & $>$28.926      & 29.007~(0.207) & 26.957~(0.077) & 26.230~(0.260) \\
S18  & 3$^{\sf h}$22$^{\sf m}$05\fs35086 & $-$15\arcdeg 23\arcmin 15\farcs2402 & 27.831~(0.143) & 27.819~(0.069) & 26.969~(0.077) & 25.581~(0.145) \\
S19  & 3$^{\sf h}$22$^{\sf m}$05\fs39387 & $-$15\arcdeg 23\arcmin 14\farcs2018 & 28.670~(0.302) & 28.417~(0.118) & 26.616~(0.056) & 25.072~(0.094) \\
S20  & 3$^{\sf h}$22$^{\sf m}$05\fs37179 & $-$15\arcdeg 23\arcmin 15\farcs6186 & $>$28.814      & 28.643~(0.147) & 26.660~(0.059) & 24.854~(0.075) \\
S21  & 3$^{\sf h}$22$^{\sf m}$05\fs34566 & $-$15\arcdeg 23\arcmin 15\farcs4618 & $>$28.813      & $>$29.632      & 27.291~(0.102) & 25.572~(0.138) \\
S22  & 3$^{\sf h}$22$^{\sf m}$05\fs39179 & $-$15\arcdeg 23\arcmin 14\farcs0298 & $>$28.809      & 29.135~(0.235) & 27.219~(0.100) & 25.685~(0.161) \\
S23  & 3$^{\sf h}$22$^{\sf m}$05\fs43718 & $-$15\arcdeg 23\arcmin 15\farcs6194 & 28.340~(0.241) & 29.474~(0.331) & $>$28.665      & $>$26.633 \\
S24  & 3$^{\sf h}$22$^{\sf m}$05\fs35072 & $-$15\arcdeg 23\arcmin 15\farcs6022 & $>$28.835      & 28.097~(0.089) & 26.975~(0.078) & 25.936~(0.202) \\
S25  & 3$^{\sf h}$22$^{\sf m}$05\fs38396 & $-$15\arcdeg 23\arcmin 13\farcs9958 & 28.367~(0.230) & 28.382~(0.119) & 26.046~(0.035) & 24.549~(0.057) \\
S26  & 3$^{\sf h}$22$^{\sf m}$05\fs33025 & $-$15\arcdeg 23\arcmin 15\farcs0438 & 27.518~(0.109) & 27.466~(0.055) & 26.284~(0.048) & 25.134~(0.094) \\
S27  & 3$^{\sf h}$22$^{\sf m}$05\fs46232 & $-$15\arcdeg 23\arcmin 14\farcs9126 & $>$28.841      & 28.581~(0.137) & 27.093~(0.084) & 25.445~(0.126) \\
S28  & 3$^{\sf h}$22$^{\sf m}$05\fs39871 & $-$15\arcdeg 23\arcmin 13\farcs9614 & $>$28.808      & 28.522~(0.132) & 27.052~(0.084) & 25.443~(0.137) \\
S29  & 3$^{\sf h}$22$^{\sf m}$05\fs39210 & $-$15\arcdeg 23\arcmin 13\farcs9558 & $>$28.908      & 28.386~(0.117) & 28.395~(0.288) & 26.075~(0.241) \\
S30  & 3$^{\sf h}$22$^{\sf m}$05\fs40792 & $-$15\arcdeg 23\arcmin 13\farcs8970 & $>$28.952      & 28.692~(0.154) & 27.315~(0.102) & 25.459~(0.135) \\
S31  & 3$^{\sf h}$22$^{\sf m}$05\fs32555 & $-$15\arcdeg 23\arcmin 14\farcs6310 & $>$28.961      & $>$29.657      & 28.039~(0.201) & 25.475~(0.127) \\
S32  & 3$^{\sf h}$22$^{\sf m}$05\fs32419 & $-$15\arcdeg 23\arcmin 15\farcs2534 & 27.612~(0.122) & 27.707~(0.062) & 27.701~(0.150) & $>$26.591 \\
S33  & 3$^{\sf h}$22$^{\sf m}$05\fs37008 & $-$15\arcdeg 23\arcmin 15\farcs9726 & 28.700~(0.320) & 28.055~(0.088) & 28.035~(0.206) & $>$26.599 \\
S34  & 3$^{\sf h}$22$^{\sf m}$05\fs42642 & $-$15\arcdeg 23\arcmin 15\farcs9482 & 27.164~(0.086) & 27.302~(0.045) & 26.856~(0.072) & $>$26.520 \\
S35  & 3$^{\sf h}$22$^{\sf m}$05\fs38581 & $-$15\arcdeg 23\arcmin 16\farcs0310 & 28.710~(0.329) & 27.696~(0.062) & 28.152~(0.227) & $>$26.624 \\
S36  & 3$^{\sf h}$22$^{\sf m}$05\fs34751 & $-$15\arcdeg 23\arcmin 15\farcs8186 & $>$28.843      & $>$29.672      & 27.447~(0.120) & 25.444~(0.127) \\
S37  & 3$^{\sf h}$22$^{\sf m}$05\fs47012 & $-$15\arcdeg 23\arcmin 15\farcs2990 & 26.530~(0.051) & 26.607~(0.024) & 26.564~(0.053) & 25.154~(0.095) \\
S38  & 3$^{\sf h}$22$^{\sf m}$05\fs35077 & $-$15\arcdeg 23\arcmin 13\farcs9774 & $>$28.916      & $>$29.672      & 27.501~(0.123) & 25.776~(0.173) \\
S39  & 3$^{\sf h}$22$^{\sf m}$05\fs38399 & $-$15\arcdeg 23\arcmin 13\farcs7882 & $>$28.831      & 27.847~(0.069) & 27.555~(0.138) & 25.455~(0.123) \\
S40  & 3$^{\sf h}$22$^{\sf m}$05\fs42445 & $-$15\arcdeg 23\arcmin 16\farcs0202 & $>$28.775      & 27.545~(0.057) & 26.175~(0.040) & 24.407~(0.050) \\
S41  & 3$^{\sf h}$22$^{\sf m}$05\fs33343 & $-$15\arcdeg 23\arcmin 14\farcs1682 & 26.522~(0.049) & 26.491~(0.022) & 26.209~(0.041) & 25.470~(0.128) \\
S42  & 3$^{\sf h}$22$^{\sf m}$05\fs31449 & $-$15\arcdeg 23\arcmin 15\farcs0958 & 28.411~(0.246) & 29.178~(0.236) & $>$28.665      & $>$26.583 \\
\hline
\end{tabular}
\end{table*}

\renewcommand{\arraystretch}{1.1}
\begin{table*}
\centering
\caption{\textbf{Photometric variability of the SN~2012Z progenitor system
  S1.} The 1$\sigma$ photometric uncertainties are given in
  parentheses. No corrections for Galactic or host extinction are made here.
\label{tab:variability}}
\bigskip
\footnotesize
\sffamily
\begin{tabular}{ c c c c c }
\hline
Date (UT) &  MJD    &  Exposure (sec) & Counts ($e^-$) & F555W (mag) \\
\hline
2005-08-06 & 53588.7 & 2400 &  171~(71) & 28.36~(0.45)\\
2005-08-17 & 53600.0 & 2400 &  656~(75) & 26.94~(0.13)\\
2005-08-24 & 53606.8 & 2400 &  282~(67) & 27.79~(0.26)\\
2005-08-27 & 53610.0 & 2400 &  392~(64) & 27.49~(0.18)\\
2005-09-02 & 53615.6 & 2400 &  389~(69) & 27.48~(0.19)\\
2005-09-03 & 53617.0 & 2400 &  308~(66) & 27.77~(0.23)\\
2005-09-05 & 53618.6 & 2400 &  392~(71) & 27.49~(0.20)\\
2005-09-07 & 53621.0 & 2400 &  329~(66) & 27.66~(0.22)\\
2005-09-11 & 53624.0 & 2400 &  429~(68) & 27.37~(0.17)\\
2005-09-16 & 53629.8 & 2400 &  242~(67) & 27.98~(0.30)\\
2005-09-20 & 53633.4 & 2400 &  327~(64) & 27.70~(0.21)\\
2005-09-27 & 53640.5 & 2400 &  193~(66) & 28.25~(0.37)\\
2006-10-24 & 54032.6 & 2080 &  388~(76) & 27.26~(0.21)\\
2006-10-07 & 54015.3 & 2080 &  264~(75) & 27.62~(0.31)\\
\hline
\end{tabular}
\end{table*}

\end{document}